\documentstyle[amssymb,epsfig,graphicx]{aipproc}

\begin{document}
\rightmark{\footnotesize{\hspace{1.6in} Proc. of M$^{2}$S-HTSC-VI Conference, Houston, 2000 (Physica C)}}

\title{Non-quasiparticle microwave absorption in $Bi_{2}Sr_{2}CaCu_{2}O_{8+\delta }$
\thanks{Work supported by NSF-ECS, ONR and AFOSR.}}

\author{S. Sridhar and Z. Zhai}
\address{Physics Department, Northeastern University, 360 Huntington Avenue, Boston, MA 02115}

\maketitle
\begin{abstract}
We show that a non-quasiparticle charge collective mode, in parallel and
coincident with the $d$-wave pair conductivity, leads to a quantitative
understanding of microwave surface impedance measurements on superconducting 
$Bi_{2}Sr_{2}CaCu_{2}O_{8+\delta }$. The analysis suggests an inhomogeneous
charge ground state in $Bi_{2}Sr_{2}CaCu_{2}O_{8+\delta }$ and other HTS.
\end{abstract}

\maketitle

A complete understanding of the microwave response of high $T_{c}$
superconductors has proved to be as elusive as the mechanism of
superconductivity itself. Two principal outstanding issues are:

(i) The basic mechanism of linear microwave absorption is still not
understood. While one would have thought that an identification of the order
parameter symmetry as $d_{x^{2}-y^{2}}$ would have provided quantitative
description of the linear microwave properties, this has not turned out to
be the case. Calculations of the absorption based upon $d$-wave (or even
mixed $s+d$ symmetry) are actually lower by orders of magnitude (as we show
below), even though they can account for the penetration depth measurements 
\cite{Jacobs95,Srikanth98}. A unified picture of the microwave loss of
single crystals, thin films and ceramics has not yet been achieved.

(ii) The HTS display a surprisingly high level of nonlinear response which
is not properly understood, despite several attempts at modelling it. The
measured nonlinear response (which is a major limitation of the use of the
cuprate superconductors in microwave applications) is significantly higher
than estimates based upon $d$-wave calculations \cite{Dahm98}. Particularly
intriguing manifestations of nonlinearity are the Josephson-like response in
single crystals \cite{Zhai97}, the so-called magnetic recovery effect\cite
{Choudhury95}(a) and $2$nd harmonic generation\cite{Choudhury95}(b).

In this paper we discuss a new analysis of the microwave response in HTS. In
the superconducting state, electromagnetic relaxation is shown to occur
predominantly via a charge mode even at frequencies in the $GHz$ ranges.
This model quantitatively describes the available microwave data for $%
Bi_{2}Sr_{2}CaCu_{2}O_{8+\delta }$ ($Bi:2212$) and also for several other
cuprate HTS.

\begin{figure}[tbph]
\begin{center}
\includegraphics*[width=0.45\textwidth]{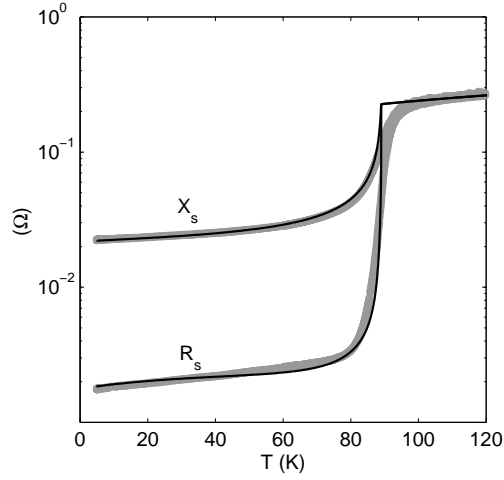}
\end{center}
\par

\caption{$R_{s}$ and $X_{s}$ {vs.} $T$ for $Bi:2212$ single crystal ( data:
thick gray line). The thin solid line represents the calculations using Eq. \ref{newzs} and \ref{newsigma}.}
\label{rsxsfig}
\end{figure}%

\begin{figure}[tbh]
  \par
\begin{center}
\includegraphics*[width=0.45\textwidth]{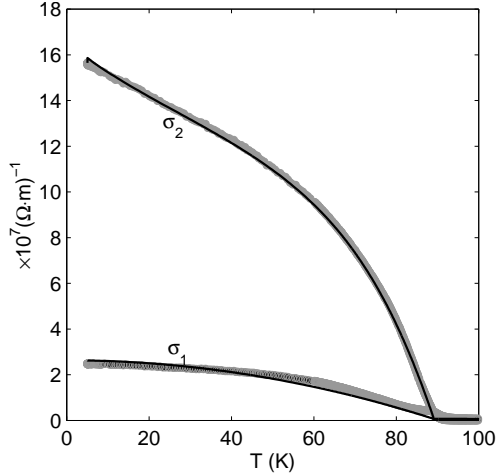}
\end{center}
\par

\caption{$\sigma _{1}$ and $\sigma _{2}$ {vs.} $T$ (data: thick gray line).
The thin solid line represents the calculations of Re[$\tilde{\sigma} _{total} $] and Im[$\tilde{\sigma}_{total}$] using Eq. \ref{newsigma}.}
\label{sigdata}
\end{figure}

Microwave data for the surface impedance $Z_{s}=R_{s}-iX_{s}$ are presented
in Fig.\ref{rsxsfig} (thick gray line). The data are obtained from $Nb$
superconducting cavity resonator measurements, which is operated in $%
TE_{011} $ mode at $10GHz$, with $H_{\omega }\,||\,\hat{c}$. The surface
impedance $Z_{s}$ and the conductivity $\tilde{\sigma}=\sigma _{1}+i\sigma
_{2}$ are related by $Z_{s}=\sqrt{-i\mu _{0}\omega /\tilde{\sigma}}$. $%
\sigma _{1}+i\sigma _{2}=-i\mu _{0}\omega /Z_{s}^{2}$ extracted from the $%
Z_{s}$ data are shown in Fig.\ref{sigdata} (thick gray line). $\sigma _{2}$
rises from zero at and above $T_{c}$ to a large value at low $T$. In the
conventional picture of microwave superconductivity, $\sigma _{2}$ is
regarded as a measure of the superfluid density, since in the standard
Mattis-Bardeen picture, $\sigma _{2}(T)=[\mu _{0}\omega \lambda
^{2}(T)]^{-1}=n_{s}(T)e^{2}/\omega m$.

In ref.\cite{Jacobs95} the $T$-dependence of $\sigma _{2}$ was shown to be
consistent with $d$-wave calculations \cite{Jacobs95}, assuming a $BCS$
temperature dependence for the $d_{x^{2}-y^{2}}$ gap parameter $\Delta
(T,\phi )=\Delta _{d}(T)\cos (2\phi )$. The d-wave model correctly describes
the low $T$ behavior of $\lambda (T)$ $\varpropto $ $T$, a feature which is
taken to be the hallmark of $d$-wave superconductivity \cite{Hardy93}(a).
Similar data and conclusions were subsequently presented in ref.\cite{Lee96}
and \cite{maedabsco}.

\begin{figure}[tbh]
 \par
\begin{center}
\includegraphics*[width=0.45\textwidth]{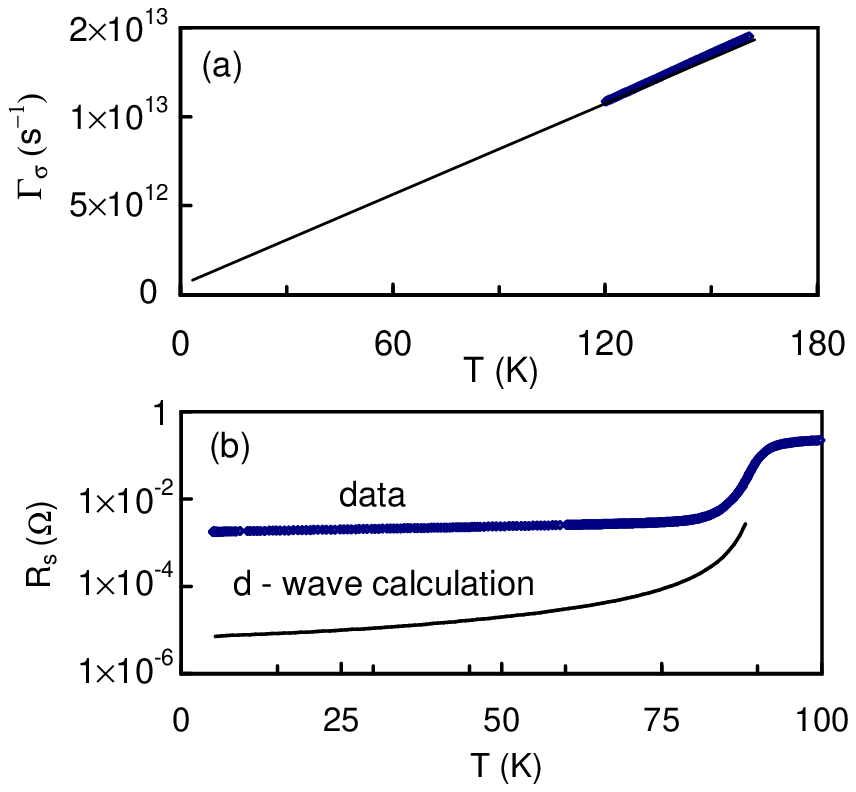}
\end{center}
\par
\caption{(a) Scattering rate $\Gamma _{\sigma }(T<T_{c})\thickapprox 9\times
10^{10}T$ (thin line) linearly extrapolated from normal state $\Gamma
_{\sigma }(T>T_{c})$ (thick dots). (b) Calculated $R_{s}(T<T_{c})$ using $d$ -wave model and $\Gamma _{\sigma }(T<T_{c})$. The calculated $R_{s}$ (thin
line) is $\sim $ $3\times 10^{2}$ smaller than experimental data (thick
dots).}
\label{scatrate}
\end{figure}

Despite the excellent agreement for $X_{s}$, and hence $\lambda (T)$ and $%
\sigma _{2}(T)$, the understanding of the absorptive part represented by the
surface resistance $R_{s}$ and the normal conductivity $\sigma _{1}$ has
remained elusive. This is demonstrated in Fig.\ref{scatrate}. If the surface
resistance is calculated using a $d$-wave gap and the scattering rate $%
\Gamma _{\sigma }(T)$ extrapolated from $T>T_{c}$ (Fig.\ref{scatrate} (a)),
then the expected $R_{s}$ would be $\sim 3\times 10^{2}$ times smaller than
that measured (Fig.\ref{scatrate} (b)). The calculation method is described
in ref.\cite{Srikanth98}. Conversely, in order to explain the large $R_{s}$
the scattering rate $\Gamma _{\sigma }(T)$ would have to drop abruptly at $%
T_{c}$ by about the order of $10^{3}$. This highly unusual behavior would
require theoretical explanation which is lacking.

Also the behavior of $\sigma _{1}$ is very anomalous. As is evident from Fig.%
\ref{sigdata} (thick gray line), $\sigma _{1}$ rises monotonically from very
low values, and shows no indication of a downturn. At low $T$, $\sigma
_{1}(4K)$ is much larger than the residual conductivity $\sigma _{res}\sim
10^{3-4}(\Omega \cdot m)^{-1}$ predicted by quasiparticle localization.
Surprisingly, essentially similar behavior of $\sigma _{1}$ in $BSCCO$ thin
films is observed at $THz$ frequencies \cite{Orenstein98}. This is
completely different from that theoretically expected for any
superconductor, where ultimately one expects that the presence of a gap
would lead to freeze out of quasiparticles. Indeed a peak is a prominent
feature of $\sigma _{1}$ in other high $T_{c}$ superconductors such as $%
YBa_{2}Cu_{3}O_{7-\delta }$ $-$ the current explanation for conductivity
peaks is that this can be understood from the expression $\sigma
_{1}=n_{qp}(T)e^{2}/m\Gamma _{\sigma }(T)$, as arising from the competition
between a decreasing $\Gamma _{\sigma }(T)$ (increasing $\tau _{\sigma
}(T)=\Gamma _{\sigma }^{-1}(T)$) and a decreasing $n_{qp}(T)\,$(due to
quasiparticle freeze-out) \cite{Hardy93}(b).

These disagreements clearly indicate that an additional mechanism is likely
to be responsible for the large microwave loss in this material. We have
achieved excellent agreement with a remarkably simple model in which we
postulate that in the superconducting state, in addition to the usual
Mattis-Bardeen complex conductivity $\tilde{\sigma}=\sigma _{s1}+i\sigma
_{s2}$, a non-quasiparticle polarization contribution also appears at $T_{c}$%
.

In conventional analyses of the microwave response of metals, the
displacement current is always ignored. This holds very well in homogeneous
metals. Thus Maxwell's equation $\nabla \times \vec{H}=\vec{J}+\partial \vec{%
D}/\partial t$, is always approximated as $\nabla \times \vec{H}=\vec{J}=%
\tilde{\sigma}\vec{E}$. This only holds true if $\omega \varepsilon ^{\prime
}<<\sigma $. This assumption may not be valid in inhomogeneous metals such
as the cuprates. We have recently thoroughly analyzed the cavity measurement
technique for the general case of a conducting (or even superconducting)
dielectric\cite{Zhai00}. Inclusion of the displacement current term then
yields a modified equation for the surface impedance as: 
\begin{equation}
Z_{s}=\sqrt{-i\omega \mu _{0}/\tilde{\sigma}_{total}}  \label{newzs}
\end{equation}
where $\tilde{\sigma}_{total}\equiv \tilde{\sigma}-i\omega \tilde{\varepsilon%
}$ is the effective conductivity. For $T<T_{c}$, $\tilde{\sigma}_{total}$ is
represented as:\ 
\begin{eqnarray}
\tilde{\sigma}_{total}\; &=&\sigma _{s1}(T)+i\sigma _{s2}(T)  \nonumber \\
&&-\frac{i\omega \varepsilon _{0}\varepsilon (T)}{1-\omega ^{2}/\omega
_{0}^{2}-i\omega \tau _{\varepsilon }(T)}\text{ .}  \label{newsigma}
\end{eqnarray}
For $T>T_{c}$, $\sigma (T>T_{c})=\sigma _{10}/t$ where $t=T/T_{c}$. We use
simple forms such as the $2$-fluid expressions for the superconducting $%
\sigma _{s1}$ and $\sigma _{s2}.$ We use $\sigma _{s2}(T)=\sigma
_{20}n_{s}(t)$, where $n_{s}(t)$ is a $d$-wave superfluid density calculated
as discussed in ref.\cite{Srikanth98} using a gap ratio $\Delta
(0)/kT_{c}=2.8$, and a monotonic temperature dependence $\sigma _{s1}=\sigma
_{10}t^{2}$ for $\sigma _{s1}(T)$.

For $\tilde{\varepsilon}$ we have used a collective mode response like a CDW
which turns on at $T_{c}$, viz. $\varepsilon (T)=\varepsilon (0)(1-t^{2})$,
where $t=T/T_{c}$ ($T_{c}=89K$). In the limit $\omega \ll \omega _{0}$,
where $\omega _{0}$ is the oscillator resonant frequency, we get a pinned
CDW $\varepsilon (T)/(1-i\omega \tau _{\varepsilon }(T))$, where $\tau
_{\varepsilon }(T)\,$is the pinning relaxation time. If $\omega \gg \omega
_{0}$, the response is a Drude-like conductivity $\varepsilon (T)/(i\omega
\tau _{\varepsilon })/(1-i\omega \tau _{\sigma }(T))$, where here $\tau
_{\sigma }=1/\omega _{0}^{2}\tau _{\varepsilon }$. Because we are
considering fixed frequency data varying $T$, either of these assumptions
are possible - they only differ in sign for the real part of $\tilde{%
\varepsilon}$. Further frequency dependent measurements are needed to
distinguish between these possibilities.

We find that excellent fits to the data are obtained in either limit. The
parameters used for the fit in the limit $\omega \ll \omega _{0}$ are: $%
\sigma _{20}=2.3\times 10^{8}(\Omega \cdot m)^{-1}$, $\sigma _{10}=7.7\times
10^{5}(\Omega \cdot m)^{-1}$, $\varepsilon (0)=1.4\times 10^{8}$, and a
temperature independent scattering time $\tau _{\varepsilon }=6.2\times
10^{-12}\sec $. Thus in this case $\omega \tau _{\varepsilon }=0.39$. In the 
$\omega \gg \omega _{0}$ Drude conductivity limit, the corresponding
parameters are: $\sigma _{20}=1.6\times 10^{8}(\Omega \cdot m)^{-1}$, $%
\sigma _{10}=7.7\times 10^{5}(\Omega \cdot m)^{-1}$, $\varepsilon (0)/\omega
\tau _{\varepsilon }=4.8\times 10^{7}$, and a temperature independent
scattering time $\tau _{\sigma }=2.2\times 10^{-13}\sec $. Clearly the model
describes the data extremely well (see Fig.\ref{rsxsfig} and \ref{sigdata}
solid thin lines). The model shows that although the reactive response is
dominated by the pair conductivity $\sigma _{2}$, the absorptive part is
completely determined by the non-quasiparticle channel and not by
quasiparticles.

A key outcome is that this approach yields an alternative explanation of the
``conductivity'' peaks mentioned previously. We now identify these peaks as
relaxation loss peaks which occur at a peak temperature $T_{p}$ where $%
\omega \tau _{\varepsilon }(T_{p})=1$. The peak corresponds to a crossover
from $\omega \tau _{\varepsilon }\ll 1$ at high $T$ to a regime $\omega \tau
\gg 1$ at low $T$. We have observed such peaks in non-superconducting
oxides, such as $Sr_{14}Cu_{24}O_{41}$ and insulating $YBa_{2}Cu_{3}O_{6.0}$ 
\cite{Zhai00,Zhai00b}. In $Bi:2212$ the peak is not observed because $\omega
\tau _{\varepsilon }(T)<1$ at all temperatures. In contrast in almost all
the other HTS, the scattering rate $\Gamma _{\varepsilon }(T)=\tau
_{\varepsilon }^{-1}(T)\,$is significantly smaller or $\tau _{\varepsilon
}(T)$ is larger, and the relaxation rate crosses the probe frequency even at
microwave frequencies, leading to a peak, as seen in superconducting $%
YBa_{2}Cu_{3}O_{6.95}$ \cite{Srikanth98}.

The dielectric strength $\varepsilon (0)\sim 10^{8}$ is typical of values in
low dimensional $CDW$ systems \cite{CDW}. The presence of such appreciable
polarization may be indicative of spatial modulation of charge. This is in
clear contrast with LTS, and may well be associated with the presence of
stripes in HTS \cite{stripes}. Note that the present results for $H_{\omega
}\,||\,\hat{c}$ are distinct from the c-axis polarization measured by ref.%
\cite{maeda-caxis}. It should also be noted that large dielectric constants
both in-plane and along the c-axis have been measured in the parent compound 
$Bi_{2}Sr_{2}(Dy,Er)Cu_{2}O_{8}$\cite{bsccodielec}.

The onset of strong polarization at or near $T_{c}$ may not be surprising in
view of the many reports of structural or lattice distortions reported at $%
T_{c}$ in $YBCO$ , $Hg:1201$ and $Tl:2212$\cite{Sharma}. It is quite
reasonable that these structural distortions are accompanied by charge
density instabilities resulting in large changes in polarization. It has
long been recognized that the oxide superconductors are also incipient
ferroelectrics, since ferroelectricity in perovskite oxides is well known.
Thus it is possible that the results reported here are observing strong
polarization modes associated with the $Bi-O$ and $(Ca,Sr)-O$ layers.
Theories where dielectricity and superconductivity are both present have
already been presented \cite{Shenoy99}. In $YBa_{2}Cu_{3}O_{6.02}$ we have
found evidence for dielectric transitions at $110K\,$and $55K$ \cite{Zhai00b}%
.

There are several important implications of the work presented here. The
displacement current channel, which is always neglected in analysis of
microwave response of conventional metals and superconductors, and in
previous analysis of HTS, cannot be ignored in HTS. Consequently, a charge
collective mode is essential to understand the microwave response of the
cuprate superconductors. The charge mode dynamics manifests itself via the
presence of relaxation loss peaks, which are the correct explanation of the
microwave absorption peaks even in the superconducting state.

The Mattis-Bardeen conductivity, so successful in describing the
electrodynamics of low $T_{c}$ superconductors, is clearly inadequate for
HTS. Quasiparticle contributions alone are clearly insufficient to describe
the microwave absorption, and do not account for the most prominent feature
of the microwave response of HTS $-$ the ``conductivity'' peaks which we
instead show here are more appropriately called microwave absorption peaks
or relaxation loss peaks.

The analysis presented here is not restricted to $Bi:2212$. Indeed, for the
first time, by including non-quasiparticle contributions, we are able to
quantitatively describe the microwave response of other cuprate
superconductors also \cite{Zhai00b}.

\end{document}